# Lines under the forest

Amelia Carolina Sparavigna
Dipartimento di Fisica,
Politecnico di Torino, Torino, Italy

Earthworks appear in the satellite images of deforested sections of the Amazon Basin. These images, and several other archaeological evidences, are showing that a lost civilization is waiting for further researches.

Analysing the images coming from the satellites launched for geophysical researches and environmental control, it is possible to gain some astonishing information about lost civilizations. It is the case of the "geoglyphs" discovered using the satellite imagery of deforested sections of the Amazon Basin, in the eastern Acre State, Brazil. These "geoglyphs" are earthworks that appear as squared or round structures, often connected by lines.
Built long before Christopher Columbus arrival in the new world, these earthworks are the remains of a lost civilization, according to a study published in the journal Antiquity [1,2]. The paper on Antiquity is summarizing a research started in 2001 [3,4], research aided by the use of Google Maps [5]. Some geoglyphs were discovered during fieldwork in the late 1970s, but most of the more recent discoveries have occurred by means of the Web-based program, which possesses some high-resolution satellite images, that "facilitates the search for landscape expressions of past civilizations that we call geoglyphs". Carbon-14 dating of charcoal associated with these geoglyphs indicates ages between 2500 and 1000 years B.P. [5].
According to Ref.2, "the combination of land cleared of its rainforest for grazing and satellite survey have revealed a sophisticated pre-Columbian monument-building society in the upper Amazon Basin on the east side of the Andes." The satellite inspection revealed to the archaeologists that an unknown civilization built several earthworks having a precise geometric plan, connected by straight orthogonal roads. The article tells that the "geoglyph culture" stretched over a region more than 250km across floodplains and the uplands. The geoglyphs are enormous structures, with diameters ranging from 100 to 300 meters. The structures are formed by ditches about 36 feet wide and several feet deep, lined by earthen banks up to 3 feet high, located on hills above the river valleys. The researchers argue that these structures were used for ceremonies. Based on the amount of manpower necessary to create them, it is estimated that at least 60,000-90,000 people might have lived in the area where about 300 geoglyphs have been so far localized [1,6].
In fact, the Spanish explorer Francisco de Orellana, a 16th century explorer, who was the first European to traverse the Amazon River, reported of densely populated regions. The recent discovery of the geoglyphs in Acre State, is telling that Orellana had not exaggerated the level of development among the people in Amazonia [7,8]. It is possible to argue that a large part of the local population was settled before the arrival of Europeans and that "became nomadic after the demographic collapse of the 16th and 17th century, due to European-introduced diseases", as discussed in the Wikipedia item on the "Terra Preta" [8].
"Terra Preta" is another quite interesting archaeological object found in Amazonia. The name means "black earth" in Portuguese: it is a type of dark, fertile anthropogenic soil found in the Amazon Basin, as in several other places on Earth. This soil has a very high charcoal content and was made by adding a mixture of charcoal, bones and manure to the Amazonian soil. It seems that this soil had been used by thousands of years [9]. The zones with this artificial soil are generally surrounded by the original soil that had benn not altered by the human activity. Terra Preta soils are then of pre-Columbian nature and were created by humans between 450 BC and AD 950 [9-11].
It is from the 1970s that some researchers started accumulating data on the existence of an elaborate civilization that prospered along the Amazon's rivers [5]. In the 1990s, scientists began documenting sites along the southern edge of the Amazonia [6]: one of them, the anthropologist Michael Heckenberger, worked near the Xingu River, several hundred miles east of Acre. His team found communities housing at least 50,000 people, living in what he defined as "garden cities", made up of small or medium villages built around central plazas and connected by a road system, mainly aligned with the cardinal directions (this alignment is observed also in the Acre geoglyphs, see Fig.1). Large defensive earthworks were created to protect the settlements. These villages were included in a complex system of forest and wetland management. Clark Erickson and his team have found evidence of an extensive ancient land management in Bolivia too. The landscape is covered by canals, raised fields and ring-shaped ditches and massive earthworks for trapping fish [12]. About the raised-field, a discussion is proposed at Ref.13. Raised fields can be seen in the area of geoglyphs in Brazil too (see Fig.2).

Geographically, these regions in Brazil and Colombia are contiguous and seem to have common characteristics. According to Ref.6, Heckenberger is saying that the researches evidenced many structures with a common design, built in Acre, Bolivia and near Xingu river, and this fact clearly displays that a much denser population than previously assumed was living in this area [14].

It seems then that a complete survey of the region is necessary to have a precise evaluation of the lost civilizations of Brazil and Bolivia. For the exploration of those areas still covered by the forest, a LIDAR inspection could be quite useful. In Ref.15, it was discussed how a small crater, the Whitecourt Meteorite Impact Crater, can be seen by the LIDAR, which is able to depict its shape removing the mask of the vegetation hiding it [16].

In fact LIDAR does an excellent analysis in the case of earthworks [17,18]. As discussed in Ref.17, where the researchers are mainly introducing the subject of LIDAR survey, this technology can be applied to any large prehistoric earthwork complexes, and, as they note, the inspection is quick and accurate. One of the images from Ref.17 is quite interesting: it is showing a profile of the High Bank circle-octagon complex. This structure is similar to those observed in Brazil (see Fig.3).

It could be interesting to understand the spread of the earthwork civilization across Americas and have a complete portrait of it. Let me conclude as Romain and Burk are telling: "many of earthwork sites have hidden discoveries waiting to be revealed by modern investigators".

**References**
1. 'Astonishing' ancient Amazon civilization discovery detailed, Rossella Lorenzi, Jan 15, 2010, Discovery News, http://news.discovery.com/archaeology/astonishing-ancient-amazon-civilization-discovery-detailed.html
2. Pre-Columbian geometric earthworks in the upper Purús: a complex society in western Amazonia, M. Pärssinen, D. Schaan and A. Ranzi, Antiquity, Vol.83(322), 1084–1095, 2009
3. Registry of geoglyphs of the Amazon region, Brazil (in Portuguese), A. Ranzi and R. Aguiar, Munda, 42, 87–90, 2001
4. Geoglyphs of Amazonia—Aerial Perspective, A. Ranzi, and R. Aguiar, 56 pp., Faculdades Energia, Florianópolis, Brazil, 2004
5. Internet Software Programs Aid in Search for Amazonian Geoglyphs, A. Ranzi, Eos, Vol. 88, 21-22 May 2007, p.226, 2007
6. http://archaeologynewsnetwork.blogspot.com/2010/11/amazonian-geoglyphs.html
7. Wikipedia, Francisco de Orellana, http://en.wikipedia.org/wiki/Francisco_de_Orellana
8. Wikipedia, Terra Preta, http://en.wikipedia.org/wiki/Terra_preta#Early_theories
9. http://www.abc.net.au/rural/news/content/201102/s3136085.htm, J. Lehmann. 9000 year old Terra Preta soils.
10. The timing of Terra Preta formation in the central Amazon: new data from three sites in the central Amazon, E.G. Neves, R.N. Bartone, J.B. Petersen and M.J. Heckenberger, p. 10, 2001.

11. Prehistorically modified soils of central Amazonia: a model for sustainable agriculture in the twenty-first century, Bruno Glaser, Philos Trans R Soc Lond B Biol Sci., 2007 February 28, 362(1478). Pp. 187–196, 2007, http://www.ncbi.nlm.nih.gov/pmc/articles/PMC2311424/
12. C. Erickson, Historical ecology and future explorations, in Amazonian Dark Earths: origin, properties, and management, Chapter 23, 2003,
downloadable http://marajoara.org/files/EricksonADE.pdf
13. The Titicaca basin: a paradigmatic region for multidisciplinary studies, Amelia Carolina Sparavigna, ARXIV:1011.0391 , 2011, http://arxiv.org/ftp/arxiv/papers/1011/1011.0391.pdf
14. M.J. Heckenberger, J.C. Russell, C. Fausto, J.R. Toney, M.J. Schmidt, E. Pereira, B. Franchetto, A. Kuikuro, Pre-Columbian Urbanism, Anthropogenic Landscapes, and the Future of the Amazon, Science, 2008, Vol. 321. no. 5893, pp. 1214 - 1217, DOI: 10.1126/science.11597696.
15. Small simple impact craters, Amelia Carolina Sparavigna, 2010 http://arxiv.org/ftp/arxiv/papers/1008/1008.5011.pdf
16. R.S. Kofman, C.D.K. Herd, E.L. Walton and D.G. Froese, The late Holocene Whitecourt meteorite impact crater: a low-energy hypervelocity event. 40th Lunar and Planetary Science Conference, 2009, 1942.PDF.
17. LIDAR Analyses of Prehistoric Earthworks in Ross County, Ohio, W.F. Romain and J. Burk, http://www.ohioarchaeology.org/joomla/index.php?option=com_content&task=view&id=233&Itemid=32
18. LIDAR Imaging of the Great Hopewell Road, W.F. Romain and J. Burks, Monday, 04 February 2008, http://timothy-price.com/LiDAR%20Imaging.htm

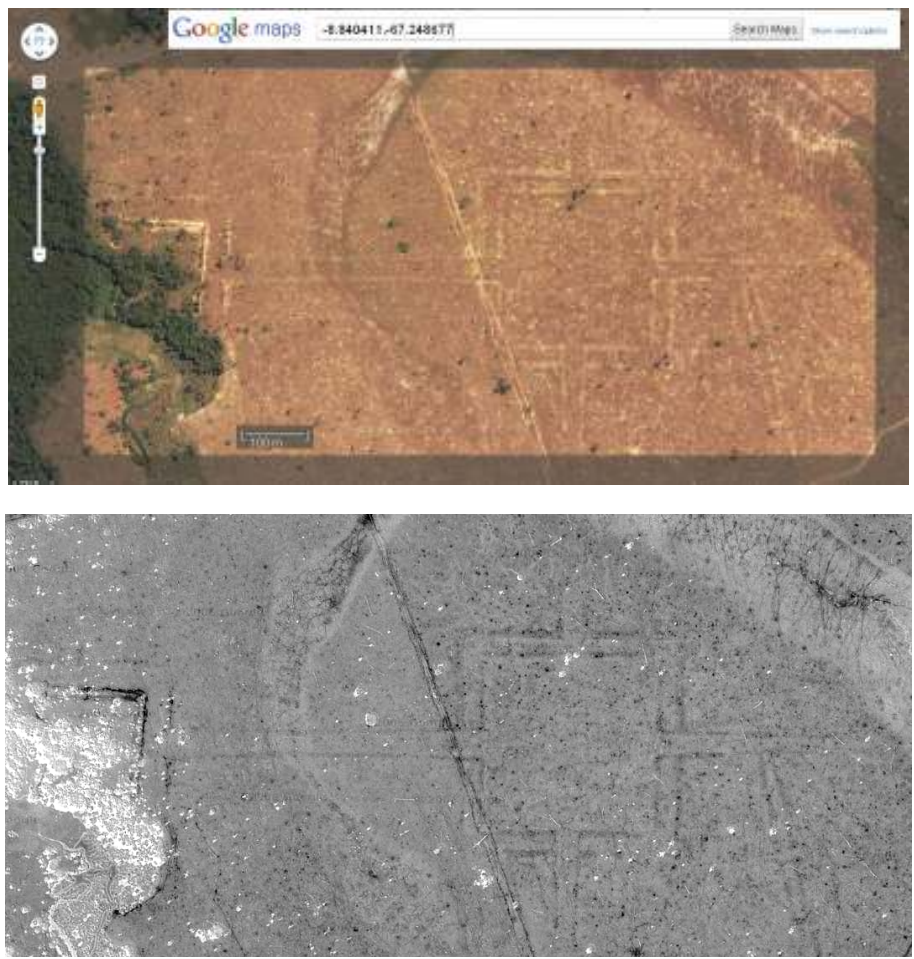

Fig.1. Near Boca do Acre, Rout BR317, two geoglyphs connected by lines. Note the almost perfect orientation of the line orientation. The original Google Maps image was enhanced with image processing. The grey-scale image is proposed for comparison.

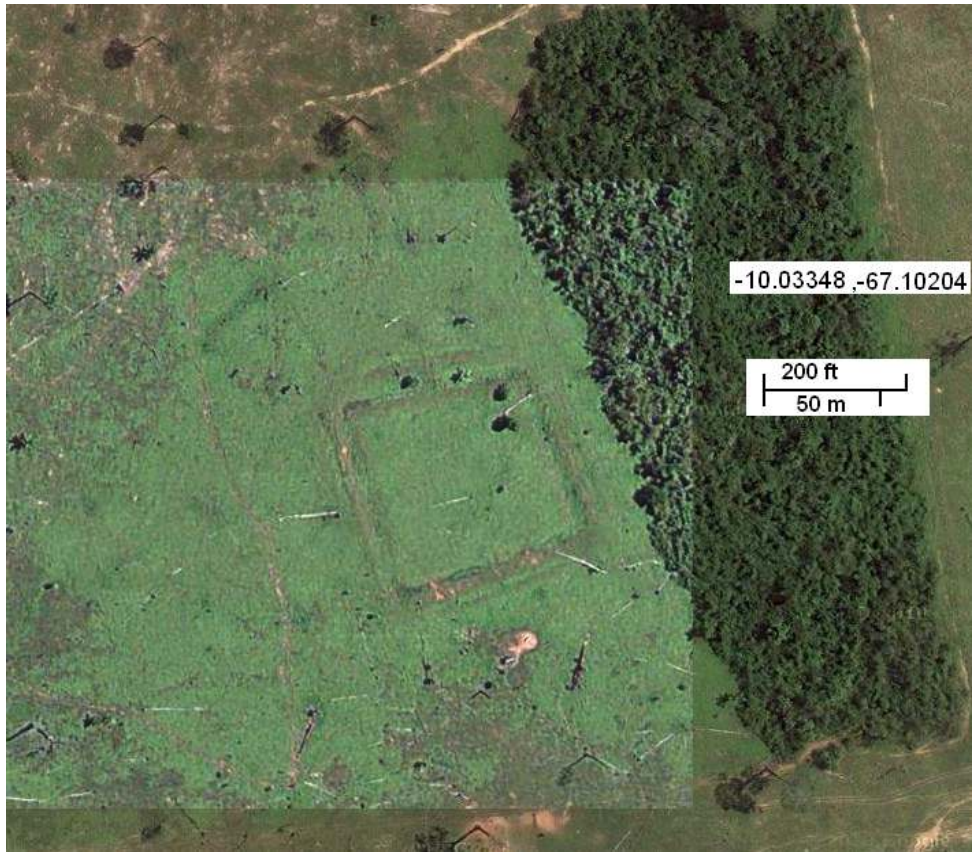

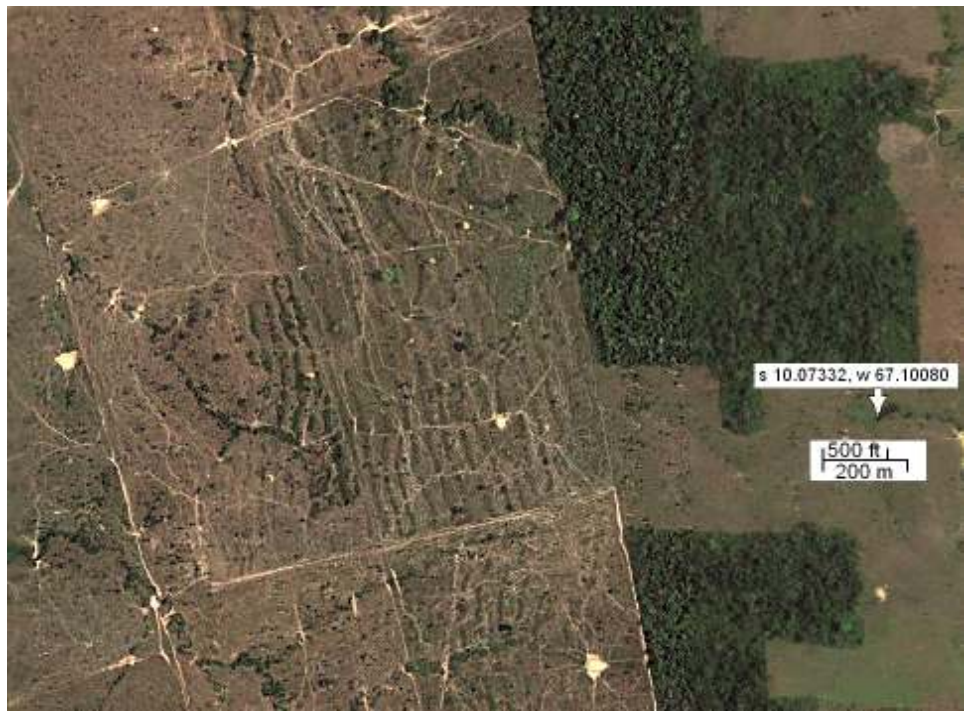

Fig.2. Near Rio Branco, Brazil, we can see by means of Google Maps, several "geoglyphs", one of them is shown in upper part of the figure. Near this geoglyphs, it is possible to observe an area with raised fields.

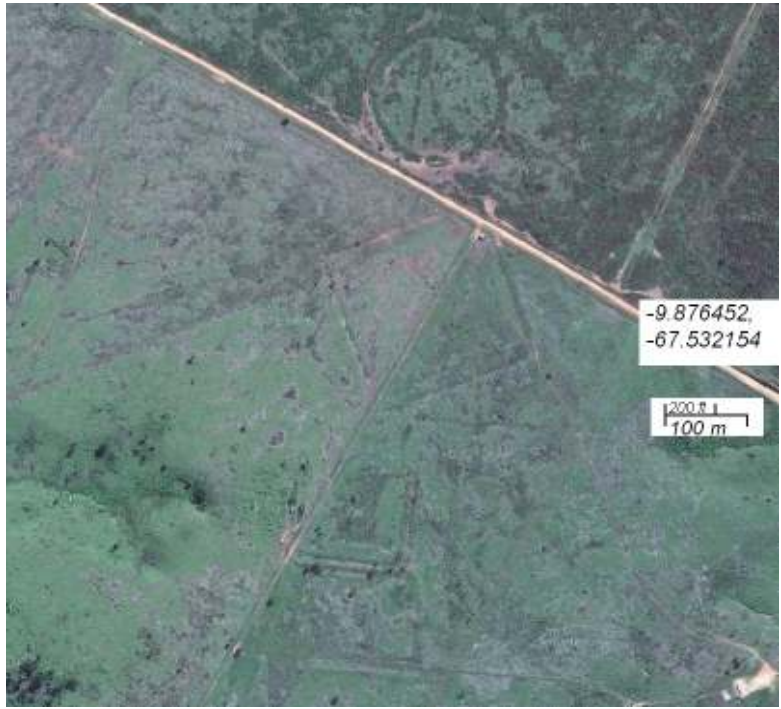

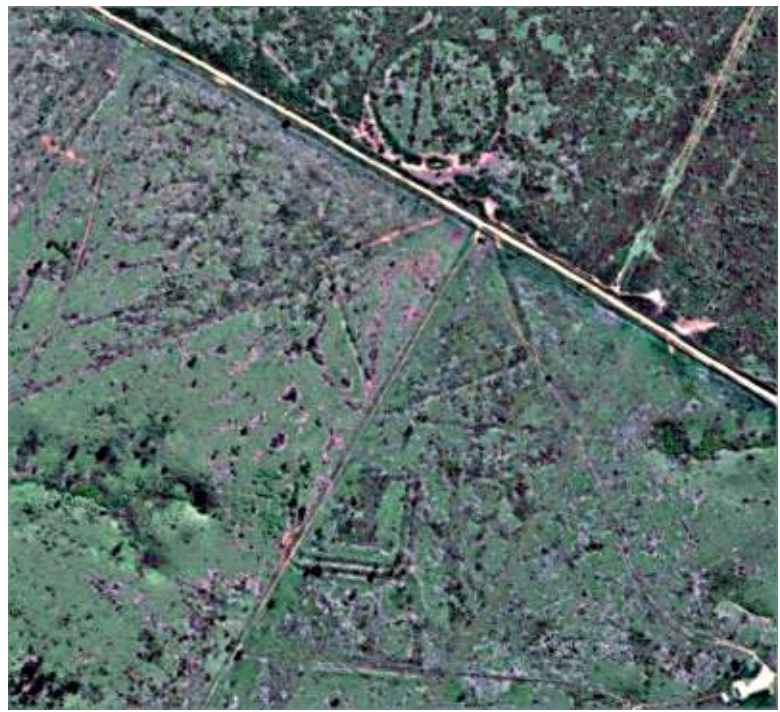

Fig.3. An earthwork complex of Rio Branco, Brazil, from Google Maps. This image is once more showing how the Google Maps imagery is fundamental in the analysis of ancient civilizations. The lower image is obtained after a processing to enhance the details.